\documentstyle[11pt]{article}
\begin{document}
  \begin{titlepage}
  \begin{flushright}
SUSSEX-AST-93/7-1\\
July 1993\\
  \end{flushright}
  \begin{center}
\vspace{1 in}

\Large
{\bf Extended Gravity Theories\\and the Einstein-Hilbert Action\\}
\vspace{.6 cm}
\normalsize
\large{David Wands} \\
\normalsize
\vspace{.6 cm}
{\em  Astronomy Centre, \\ School of Mathematical \& Physical Sciences, \\
University of Sussex, \\ Brighton BN1 9QH.\\U.~K.\\
DWANDS@STARLINK.SUSSEX.AC.UK}\\
\end{center}
\vspace{.6 cm}
\begin{abstract}
I discuss the relation between arbitrarily high-order theories of gravity
and scalar-tensor gravity at the level of the field equations and the action.
I show that $(2n+4)$-order gravity is dynamically equivalent to
Brans-Dicke gravity with an interaction potential for the Brans-Dicke
field and $n$ further scalar fields. This scalar-tensor action is then
conformally equivalent to the Einstein-Hilbert action with $n+1$ scalar fields.
This clarifies the nature and extent of the conformal equivalence between
extended gravity theories and general relativity with many scalar fields.
\end{abstract}

\begin{center}
{\small PACS number~~~~04.50}
\end{center}

\end{titlepage}
\section{Introduction}

Over recent years there has been considerable interest in gravity theories
derived from lagrangians extended beyond the Einstein-Hilbert action of
general relativity.
Originally this was motivated by the question {\em what would happen if} the
fundamental action were different in an attempt to contrast the
predictions of Einstein gravity with alternative
theories\cite{JORDAN59,BRANSDICKE,NORDVEDT70,O'HANLON72}. More recently
it has been
realised that such extensions may be inescapable if we wish to ask the
question {\em
what happens when} quantum corrections become important as they must on scales
close to the Planck length.  Stelle\cite{STELLE77} was the first to
construct a renormalisable gravity action by including terms quadratic
in the Riemmann curvature tensor.
Such terms invariably appear as
counter-terms in renormalisable theories\cite{INDUCEDGRAVITY}
involving scalar fields coupled to the curvature tensor,
suggesting that
the Einstein-Hilbert action itself is only the effective action induced by
the vacuum, as originally proposed by Sakharov\cite{SAKHAROV68}.
Any lagrangian based on a finite number of terms involving the curvature tensor
or its derivatives may be seen as a low-energy approximation to some
fundamental action.  One-loop contributions in string theory, for instance,
give a lagrangian coupling the Ricci scalar to the dilaton
field\cite{ONELOOP}.

Gravity theories where the contracted Ricci tensor appears coupled to a scalar
field I will refer to as scalar-tensor gravity.
Their study in a cosmological context has been pursued with particular vigour
in
recent years in the context of extended inflation\cite{EXTINF}.
Gravity lagrangians with terms
of quadratic or higher order in the Ricci scalar have also been studied in
cosmology\cite{BARROWOTTEWIL}
as these may be able to drive a period of inflation without the
introduction of an extra inflaton field\cite{MIJICETAL}.

Teyssandier and Tourrenc\cite{TEYTOU} pointed out that the field equations
derived
from a gravity lagrangian which is an arbitrary function of the Ricci scalar
are
identical to those from a lagrangian involving an extra degree of
freedom, a scalar field $\varphi$, which is then coupled to a linear
function of the Ricci scalar, i.~e. a scalar-tensor theory. In this
letter I will show how this can be extended to write any gravitational
lagrangian which is a function of the Ricci scalar, arbitrarily high
derivatives $\Box^nR$, and a set of non-minimally coupled scalar fields
$\phi_i$, as scalar-tensor gravity.  This then allows one to write down the
conformal transformation to the Einstein-Hilbert
action of general relativity with many scalar fields.

\section{Fourth-order gravity as scalar-tensor gravity}

The most commonly studied extended gravity action based solely on a
function of the Ricci scalar is the quadratic action
\begin{equation}
S = \frac{1}{16\pi G_D} \int_M d^Dx \sqrt{-g} \left[ R + \alpha R^2
      + 16\pi G_D{\cal L}_{\rm matter} \right]
\end{equation}
where $M$ is a manifold of $D$-dimensions.
(I follow the sign conventions of
Wald\cite{WALD84}.) One might also include terms in
$R_{\mu\nu\lambda\kappa}R^{\mu\nu\lambda\kappa}$ and
$R_{\mu\nu}R^{\mu\nu}$.
The first of these can always be eliminated as part of a total
divergence while the second can only be rewritten in terms of $R^2$ in
homogeneous and isotropic metrics.  It could be treated in a similar
manner to the $R^2$ term in what follows, but this would require the
introduction of a tensor field $\varphi_{\mu\nu}$\cite{KASPER93}
and so cannot be
described as scalar-tensor gravity. Considering the action quadratic
only in the Ricci scalar, variation with respect to the
metric $g_{\mu\nu}$ yields the Euler-Lagrange field equations
\begin{equation}
\label{QUADEOM}
(1+2\alpha R) (R_{\mu\nu} - \frac{1}{2}g_{\mu\nu}R) = 8\pi G_D T_{\mu\nu}
 - \frac{\alpha}{2}R^2g_{\mu\nu}
 + 2\alpha (g_{\mu}^{\lambda}g_{\nu}^{\kappa} - g_{\mu\nu}g^{\lambda\kappa})
 R_{;\lambda\kappa}
\end{equation}
Thus we find terms in the field equations which are second derivatives of the
Ricci tensor and thus fourth derivatives of the metric.

However, Teyssandier and Tourrenc\cite{TEYTOU} pointed out that the field
equations are the same as would be derived if we considered an action linear
with respect to the Ricci scalar including a new non-dynamical field $\varphi$.
\begin{equation}
S_{\varphi} = \frac{1}{16\pi G_D} \int_M d^Dx \sqrt{-g} \left[
 (1+2\alpha\varphi) R - \alpha\varphi^2 + 16\pi G_D {\cal L}_{\rm matter}
 \right]
\end{equation}
We now have equations of motion from varying the action with respect to the
metric and the field $\varphi$.
\begin{eqnarray}
(1+2\alpha\varphi) (R_{\mu\nu} - \frac{1}{2}g_{\mu\nu}R) & = &
 8\pi G_D T_{\mu\nu}
 - \frac{\alpha}{2}\varphi^2g_{ \mu\nu}
 + 2\alpha (g_{\mu}^{\lambda}g_{\nu}^{\kappa} - g_{\mu\nu}g^{\lambda\kappa})
 \varphi_{;\lambda\kappa}\\
2\alpha ( R - \varphi ) & = & 0
\end{eqnarray}
The last equation simply requires $R=\varphi$ for $\alpha\neq0$
($\alpha=0$ corresponds to general relativity anyway) and substituting
this into the metric field equation clearly gives the original field equations
for the quadratic action, equation(\ref{QUADEOM}).  The field equations
only explicitly contain second derivatives of the metric, but do include
second derivatives of the new field $\varphi$ which is set equal
to the Ricci scalar (containing second derivatives of the metric) by its
equation of motion.

The action $S_{\varphi}$ is clearly a scalar-tensor theory, albeit one
without a kinetic term for the scalar field (equivalent to Brans-Dicke
gravity\cite{BRANSDICKE} with the Brans-Dicke parameter $\omega=0$) but with a
potential term $\propto \alpha\varphi^2$.  This is identical to
O'Hanlon's massive dilaton gravity\cite{O'HANLON72}
\begin{equation}
S_{\varphi} = \frac{1}{16\pi} \int_M d^Dx \sqrt{-g} \left[
 \Phi R - m^2 f(\Phi) + 16\pi{\cal L}_{\rm matter} \right]
\end{equation}
where $\Phi \equiv (1+2\alpha\varphi)/G_D$ and
$m^2f(\Phi)=\alpha\varphi^2/G_D=(G_D\Phi-1)^2/(4\alpha G_D)$.
O'Hanlon introduced
his action to produce a ``fifth force'' Yukawa type interaction
in the quasi-Newtonian gravitational potential,  so it is
not surprising that it also appears in the weak-field limit of
fourth-order gravity\cite{PECHLANERSEXL}.

This equivalence applies to any
gravitational lagrangian that is a function of the Ricci scalar.
\begin{eqnarray}
S & = & \frac{1}{16\pi G_D} \int_M d^Dx \sqrt{-g} \left[
 F(R) + 16\pi G_D {\cal L}_{\rm matter} \right] \\
& \rightarrow & \frac{1}{16\pi G_D} \int_M d^Dx \sqrt{-g} \left[
 F(\varphi) + (R - \varphi) F'(\varphi) + 16\pi G_D {\cal L}_{\rm matter}
 \right]
\end{eqnarray}
Where $'$ denotes differentiation with respect to $\varphi$.
The field equation for $\varphi$ again gives $\varphi = R$, provided
$F''\neq0$, which substituted into the metric field equations
gives the standard field equations for the $F(R)$ lagrangian.

In O'Hanlon's notation, we have
\begin{equation}
G_D \Phi \equiv F'(\varphi)
\hspace{1in}
G_D m^2 f(\Phi) = \varphi F' - F
\end{equation}

\section{Higher-order gravity as scalar-tensor
gravity}

We can extend this equivalence to sixth- or arbitrarily high-order
gravity theories derived from lagrangians that are functions not only of
$R$ but also $\Box R$ or $\Box^n R$, where $\Box$ is the d'Alembertian.
Such terms can also be generated by quantum corrections to general
relativity and these theories have recently been studied in the context of
inflationary cosmology\cite{SCHMIDT90,SIXTHORDER}.

Notice that apparently different
lagrangians $F(\Box^iR)$ can yield identical field equations if they
differ only by a total divergence, as this can be written as a boundary
term which cannot contribute to the Euler-Lagrange equations. For instance,
$\Box^iR$ alone is a total divergence and can be ignored, while a term
$\Box^nR\Box^mR$ can be integrated by parts to give $R\Box^{m+n}R$.
Thus I can take any polynomial function $F(\Box^iR)$ to be
linear in its highest-order derivative ($\Box^nR$), and that this term
is multiplied only by a function of the Ricci scalar, $F_n(R)$. Its field
equations contain $(2n+4)$-order derivatives of the metric tensor. Any
lagrangian containing a term $\Box^nR$ multiplied by derivatives of the
Ricci scalar, or any other fields, corresponds to a
still-higher-order gravity theory as it can
be integrated by parts to give terms at least of order $\Box^{n+1}R$.

Consider then the action for $(2n+4)$-order gravity, with a polynomial
$F(\Box^iR)=F_0(\Box^{i\neq n}R) + F_n(R)\Box^n R$
\begin{equation}
S = \frac{1}{16\pi G_D} \int_M d^Dx \sqrt{-g} \left[
 F(R,\Box R,...\Box^nR) + 16\pi G_D {\cal L}_{\rm matter} \right]
\end{equation}
The classical field equations are\cite{SCHMIDT90}
\begin{eqnarray}
\Theta \left( R^{\mu\nu} - \frac{1}{2}g^{\mu\nu} R \right)
& = & 8\pi G_D T^{\mu\nu} + \frac{1}{2}g^{\mu\nu} (F-\Theta R)
 + ( g^{\mu\lambda}g^{\nu\kappa} - g^{\mu\nu}g^{\lambda\kappa} )
  \Theta_{;\lambda\kappa} \nonumber \\
\nonumber \\
& & \, \, + \frac{1}{2} \sum_{i=1}^{n} \sum_{j=1}^{i}
  ( g^{\mu\nu}g^{\lambda\kappa} + g^{\mu\lambda}g^{\nu\kappa} )
   (\Box^{j-1}R)_{;\kappa}
   \left( \Box^{i-j} \frac{\partial F}{\partial \Box^i R} \right)_{;\lambda}
\nonumber \\
& & \, \, \, \,
  - g^{\mu\nu}g^{\lambda\kappa} \left( (\Box^{j-1}R)_{;\kappa} \Box^{i-j}
   \frac{\partial F}{\partial \Box^i R} \right)_{;\lambda}
\end{eqnarray}
where I have used
\begin{equation}
\Theta \equiv \sum_{j=0}^{n} \Box^j \left( \frac{\partial F}{\partial\Box^jR}
 \right)
\end{equation}

We introduce the new variables
${\varphi_i}=\varphi_0,\varphi_1,...\varphi_n$ and write the function
$F=F({\Box^iR})$ as $F({\varphi_i})$, selecting a new action whose field
equations will require $\varphi_i=\Box^iR$, with $\Box^0R \equiv R$.
\footnote{Notice
that these fields have the non-standard dimensions of
$({\rm length})^{-2(i+1)}$. We could re-define the fields to give them
more usual dimensions of mass for a scalar field, but for brevity of notation
I will stick to $\varphi_i$ as defined here.}
Thus we choose the dynamically equivalent action
\begin{equation}
S' = \frac{1}{16\pi G_D} \int_M d^Dx \sqrt{-g} \left[
 F({\varphi_i})
 + \sum_{j=0}^{n} \frac{\partial F({\varphi_i})}{\partial\varphi_j}
  \left( \Box^jR - \varphi_j \right)
 + 16\pi G_D {\cal L}_{\rm matter} \right]
\end{equation}
and the $\varphi_k$ field equations are
\begin{equation}
\sum_{j=0}^n F_{jk} ( \Box^jR - \varphi_j ) = 0
\end{equation}
where
\begin{equation}
\label{Fjk}
F_{jk}=\frac{\partial}{\partial\varphi_k}
 \frac{\partial F({\varphi_i})}{\partial \varphi_j}
\end{equation}
Then this does indeed require $\Box^iR=\varphi_i$, subject now to the
condition that the matrix
$F_{jk}$ is non-degenerate (i.~e. $\det F_{jk}\neq0$), and it can be
verified that the metric
field equations are indeed the same as in the original higher-order gravity
theory.

The lagrangian, which still contains derivatives of $R$, can be reduced
to a lagrangian linear in the Ricci scalar by integration by parts. For
instance
\begin{eqnarray}
\int_M d^Dx \sqrt{-g} \frac{\partial F}{\partial\varphi_i} \Box R
 & = & \int_M d^Dx \sqrt{-g} R \Box \frac{\partial F}{\partial\varphi_i}\\
 &   & \, + \int_{\partial M} d^{D-1}x \sqrt{h} n^{\mu} \left(
 \frac{\partial F}{\partial \varphi_i} \nabla_{\mu} R
 - R \nabla_{\mu} \frac{\partial F}{\partial\varphi_i} \right)
\end{eqnarray}
Because I am only concerned in this section with
actions yielding equivalent
Euler-Lagrange field equations I will continue to disregard boundary
terms. That is, I am assuming that to obtain the variation of the
action to first-order, I can neglect the contribution due to the
variation of the fields' derivatives on the boundary, $\partial M$.
There is nothing lost by this, as the correct form of the boundary term
required in higher-order gravity theories to avoid this requirement is not in
general known\cite{MADSENBARROW89};  a fact I will return to later.

Thus the action can be brought to
\begin{equation}
S_{\varphi_i} = \frac{1}{16\pi G_D} \int_M d^Dx \sqrt{-g} \left[
 \left( \sum_{j=0}^n \Box^j \frac{\partial F}{\partial \varphi_j}
  \right) R
 + F({\varphi_i})
 - \sum_{j=0}^n \varphi_j \frac{\partial F}{\partial\varphi_j}
 + 16\pi G_D {\cal L}_{\rm matter} \right]
\end{equation}
Notice that the scalar functional, $\Phi$, multiplying the Ricci scalar
will contain derivatives of the scalar fields $\varphi_i$
\begin{equation}
G_D\Phi = \sum_{j=0}^n \Box^j \frac{\partial F}{\partial \varphi_j}
\end{equation}
For this to correspond to the standard scalar-tensor lagrangian we must
consider the functional $\Phi$ as a scalar field. (This is another subtle
change in the underlying action, although again it will yield a
dynamically equivalent theory as the classical field equations are not
changed.) In practice it is easy to eliminate $\varphi_n$ by rewriting it as
a functional of $\Phi$ and the other fields $\varphi_{i\neq n}$, as it
only appears linearly in the above expression for $\Phi$ in the $j=0$ term.
\begin{equation}
\varphi_n \frac{dF_n}{d\varphi_0} = G_D\Phi
 - \sum_{j=0}^{n-1} \Box^j \frac{\partial F_0}{\partial \varphi_j}
 - \Box^n F_n
\end{equation}

This method can be straightforwardly extended to higher-order gravity
lagrangians that also contain scalar fields already coupled to the Ricci
scalar or its derivatives. Thus $(2n+4)$-order scalar-tensor gravity
with $m$ non-minimally coupled scalar fields can be written as a
standard (second-order) scalar-tensor theory with $m+n+1$ scalar fields.

\subsection{Example 1: $F=R+\gamma R\Box R$}

This corresponds to sixth-order gravity\cite{SIXTHORDER} and so I will show it
to be dynamically equivalent to second-order gravity with two scalar fields.
$F(\varphi_i)=\varphi_0(1+\gamma\varphi_1)$ and
so the equivalent action (in $4$-dimensional spacetime) is
\begin{equation}
S_{\varphi_i} = \frac{1}{16\pi G} \int_M d^4x \sqrt{-g} \left[
 \left( 1 + \gamma\varphi_1 + \gamma\Box\varphi_0
  \right) R - \gamma\varphi_0\varphi_1
 + 16\pi G {\cal L}_{\rm matter} \right]
\end{equation}
The scalar multiplying the Ricci scalar is thus
$G\Phi = 1 + \gamma\varphi_1 + \gamma\Box\varphi_0$.
The $\varphi_i$ field equations are
\begin{eqnarray}
\gamma\Box R & = & \gamma\varphi_1 \\
\gamma R & = & \gamma\varphi_0
\end{eqnarray}
So the requirement that the matrix $F_{jk}$ (equation \ref{Fjk}) is
non-degenerate is simply that $\gamma\neq0$. If we wish to treat $\Phi$
as a variable rather than $\varphi_1$ the action can be rewritten, again
neglecting boundary terms, as
\begin{equation}
S_{\Phi} = \frac{1}{16\pi G} \int_M d^4x \sqrt{-g} \left[
 G\Phi R - \varphi_0(G\Phi -1) - \gamma
g^{\mu\nu}\varphi_{0,\mu}\varphi_{0,\nu}
 + 16\pi G {\cal L}_{\rm matter} \right]
\end{equation}
Notice now that the derivative terms now appear
as a kinetic term for the field $\varphi_0$ in the action rather than
multiplying the Ricci scalar. To make this field appear as a canonical
scalar field with the usual kinetic term in the matter lagrangian we can
define
\begin{equation}
\sigma \equiv \sqrt{\frac{\gamma}{8\pi G}} \varphi_0
\end{equation}
Thus this sixth-order gravity theory can be seen
to be dynamically equivalent to Brans-Dicke theory with $\omega=0$ (as
there is no kinetic term for the Brans-Dicke field $\Phi$) and an
interaction potential
\begin{equation}
V(\Phi,\sigma) = \frac{\sigma(G\Phi-1)}{\sqrt{32\pi\gamma G}}
\end{equation}
coupling $\Phi$ to the scalar field $\sigma$.

\subsection{Example 2: $F(\phi,R) = R+\alpha R^2
 -8\pi G(\xi\phi^2R+g^{\mu\nu}\phi_{,\mu}\phi_{,\nu})$}

The field equations from this lagrangian contain terms of fourth-order
in the metric with a non-minimally coupled scalar field $\phi$ and has
been considered in the context of inflationary cosmology by
Maeda {\it et al}\cite{MAEDAETAL89}.  It can be rewritten by
introducing the dynamically equivalent action whose field equations are
second-order with respect to the metric with two scalar fields.
\begin{eqnarray}
S & = & \frac{1}{16\pi G} \int_M d^4x \sqrt{-g}
 \left[
 (1-8\pi G\xi\phi^2+2\alpha\varphi)R
 -\alpha\varphi^2 \right. \nonumber \\
& & \hspace{5cm} \left. +16\pi G \left(
 -\frac{1}{2}g^{\mu\nu}\phi_{,\mu}\phi_{,\nu} + {\cal L}_{\rm matter}
 \right)  \right]
\end{eqnarray}
This can be rewritten in terms of the single non-minimally coupled Brans-Dicke
field, $\Phi$, with no kinetic term if we define
\begin{equation}
G\Phi = 1 - 8\pi G\xi\phi^2 + 2\alpha\varphi
\end{equation}
and substitute for $\varphi$ in terms of this field:
\begin{eqnarray}
\label{EX2ACTION}
S & = & \frac{1}{16\pi} \int_M d^4x \sqrt{-g} \left[ \Phi R
 + 16\pi \left(
 -\frac{1}{64\pi\alpha G} (G\Phi -1 +8\pi G\xi\phi^2)^2
\right. \right. \nonumber \\
& & \hspace{6cm}\left. \left. \: \:
 -\frac{1}{2}g^{\mu\nu}\phi_{,\mu}\phi_{,\nu} + {\cal L}_{\rm matter}
 \right)  \right]
\end{eqnarray}

\section{Conformal equivalence}

I have shown the dynamical equivalence of between the field equations
derived from arbitrarily high-order gravity lagrangians and
scalar-tensor theories. This should not be very
surprising to those who know that both theories are
conformally related to general relativity plus scalar fields\cite{MAEDA89}.
Having conformally transformed from the fourth-order
field equations, for instance, to an Einstein metric, one introduces a scalar
field, $\psi$,
which could be interpreted as the conformal transform of a Brans-Dicke
field $\ln G_D\Phi \propto \psi$ (with $\omega=0$ and a self-interaction) and
so transform back to the scalar-tensor field equations.

It is important to stress however that the conformal equivalence between
scalar-tensor gravity and general relativity with a scalar field, in
contrast to my discussion of the relation between scalar-tensor and
higher-order gravity, is {\em not} just a dynamical equivalence. The
gravitational {\em action} is conformally equivalent, not just the
solutions to the classical equations of motion.

If we write
the action in scalar-tensor form, i.~e. in terms of what I will call the Jordan
metric, $g_{\mu\nu}$,
\begin{eqnarray}
S_{\Phi} & = & \frac{1}{16\pi} \int_M d^Dx \sqrt{-g} \left[ \Phi R
 - \frac{\omega}{\Phi} g^{\mu\nu} \Phi_{,\mu} \Phi_{,\nu} -
 2\Lambda(\Phi)
 + 16\pi {\cal L}_{\rm matter} \right]
 \nonumber \\
 & & \ \ + \frac{1}{8\pi} \int_{\partial M} d^{D-1}x \sqrt{h} \Phi K
\end{eqnarray}
I have included here a boundary term which is necessary, just as it is in
general relativity\cite{GRBOUNDARY}, if we are to derive
the field equations from the
requirement that the action is stationary with respect to first-order
variations of
the fields subject only to the constraint that the variations vanish on
the boundary, $\partial M$, of the manifold. $h_{\mu\nu}$ is the metric
on this $(D-1)$-dimensional surface and $K$ its extrinsic curvature.
The field equations derived are then the usual equations of motion for
scalar-tensor gravity.

One can always write the field
equations {\em and} the action in terms of a conformally rescaled metric
\begin{equation}
\tilde{g}_{\mu\nu} = \Omega^2 g_{\mu\nu}
\end{equation}
where the conformal factor $\Omega^2=(G_D\Phi)^{2/(D-2)}$.
This is just a change of
variables which is always possible for $\Phi>0$, equivalent to
demanding that the gravitational coupling be positive definite which may
well be required anyway.
Written in terms of this new metric the action becomes
\begin{eqnarray}
\label{E-H}
S_{\Phi} & = & \frac{1}{16\pi G_D} \int_M d^Dx
 \sqrt{-\tilde{g}} \left[  \tilde{R}
 + 16\pi G_D \left( -\frac{1}{2} \tilde{g}^{\mu\nu} \psi_{,\mu} \psi_{,\nu}
 \right. \right. \nonumber \\
& & \hspace{7cm} \left. \left.
 - V(\psi) + (G_D\Phi)^{-D/(D-2)} {\cal L}_{\rm matter} \right) \right]
  \nonumber \\
& & \: \:
 + \frac{1}{8\pi G_D} \int_{\partial M} d^{D-1}x \sqrt{\tilde{h}} \tilde{K}
\end{eqnarray}
where I have been careful not to discard any boundary terms. This is the
full general relativistic action, plus a
scalar field $\psi$ defined by
\begin{equation}
d\psi = \sqrt{\frac{\frac{D-1}{D-2}+\omega(\Phi)}{8\pi G_D}} \frac{d\Phi}{\Phi}
\end{equation}
whose potential is
\begin{equation}
V(\psi) = \frac{\Lambda}{8\pi G_D} (G_D\Phi)^{-D/(D-2)}
\end{equation}

The difference between the two frames is in the form of the matter
lagrangian. If some fluid is defined in the Jordan metric, then in the
Einstein frame the factor $\Omega^{-D}=(G_D\Phi)^{-D/(D-2)}$
in the matter lagrangian
introduces an interaction between this fluid and the scalar
field $\psi$. It is the nature of the matter lagrangian which may be
used to specify the ``physical'' metric\cite{BRANS88}.

For higher-order gravity theories the conformal transform is not
so straightforward. In the case of $F(R)$ lagrangians, for instance, the
conformal factor is $\Omega^2=(dF/dR)^{2/(D-2)}$\cite{MAEDA89,BARROWCOTSAKIS88}
and the scalar field that appears in the Einstein frame is
\begin{equation}
\psi = \sqrt{\frac{D-1}{8\pi (D-2)G_D}} \ln \frac{dF}{dR}
\end{equation}
This is a new variable in the Einstein frame but does not appear to
correspond to any extra degree of freedom in the higher-order theory,
but rather a function of the Ricci scalar for the metric. In fact the Einstein
frame described as the conformal transform of higher-order gravity
theories is actually the conformal transformation of the dynamically
equivalent scalar-tensor theory described earlier. The new degree of freedom
$\psi$ is just the variable $\varphi$, which along classical
trajectories obeys $\varphi = R$.

I have shown that the more general case of arbitrarily high-order gravity
based on lagrangians
that are functions of $R,\Box R,...,\Box^nR$, are also dynamically equivalent
to
scalar-tensor gravity and again it is this lagrangian, given in terms of
the scalar fields $\Phi$ and $\varphi_i$, for $i=0,1,...(n-1)$, that is
conformally equivalent to general relativity plus $n+1$ scalar fields.
The original higher-order gravity theory is only equivalent to the
conformally related Einstein theory at the level of the classical field
equations, while the Einstein action is actually equal to that in the
scalar-tensor theory, not only along the classical trajectories.

In this case the conformal factor\cite{SCHMIDT90}
\begin{equation}
\Omega^2 = (G_D\Phi)^{2/(D-2)} =
 \left( \sum_{j=0}^n \Box^j \frac{\partial F}{\partial\varphi_j}
  \right)^{2/(D-2)}
\end{equation}
In general we can only define one scalar field in the Einstein
frame which has a standard kinetic term:
\begin{equation}
\psi = \sqrt{\frac{D-1}{8\pi (D-2)G_D}} \ln G_D\Phi
\end{equation}
The remaining scalar degrees of freedom in the
Einstein frame have non-standard kinetic terms and may also be coupled via
the potential
\begin{equation}
V = - \frac{F(\varphi_i) + \sum_{j=0}^{n} \varphi_j
 \frac{\partial F}{\partial\varphi_j}}{16\pi G_D(G_D\Phi)^{D/(D-2)}}
\end{equation}

\subsection{Example 1: $F=R + \gamma R\Box R$}

The dynamically equivalent scalar-tensor theory to this sixth-order
gravity lagrangian was shown earlier to have the full action
\begin{eqnarray}
S_{\Phi} & = & \frac{1}{16\pi G} \int_M d^4x \sqrt{-g} \left[
 G\Phi R + 16\pi G \left( - \frac{\sigma(G\Phi -1)}{\sqrt{32\pi\gamma G}}
 - \frac{1}{2}g^{\mu\nu}\sigma_{,\mu}\sigma_{,\nu}
 + {\cal L}_{\rm matter} \right) \right] \nonumber \\
&   & \ \ + \frac{1}{8\pi} \int_{\partial M} d^3x
 \sqrt{h} \Phi K
\end{eqnarray}
with the two scalar fields
obeying $\sigma=(\sqrt{\gamma/8\pi G})R$ and $G\Phi = 1+2\gamma\Box R$
along classical
solutions. Rewritten in terms of the conformally rescaled Einstein
metric, $\tilde{g}_{\mu\nu}=(G\Phi) g_{\mu\nu}$, this becomes the
Einstein-Hilbert action (equation \ref{E-H}) with the new matter lagrangian
\begin{eqnarray}
\tilde{\cal L}_{\rm matter} & = & -\frac{1}{2}\tilde{g}^{\mu\nu}
 \left( \psi_{,\mu} \psi_{,\nu}
  + \exp(-\sqrt{16\pi G/3}\psi)
   \sigma_{,\mu} \sigma_{,\nu} \right) \nonumber \\
 & & \ \ - V (\psi,\sigma) + \exp(-2\sqrt{16\pi G/3}\psi) {\cal L}_{\rm matter}
\end{eqnarray}
where $\psi = \sqrt{\frac{3}{16\pi G}} \ln G\Phi$ and the potential
\begin{equation}
V (\psi,\varphi_0) = \frac{\sigma}{\sqrt{32\pi\gamma G}}
 \left( \exp\left(-\sqrt{\frac{16\pi G}{3}}\psi\right)
 - \exp\left(-2\sqrt{\frac{16\pi G}{3}}\psi\right) \right)
\end{equation}

\subsection{Example 2: $F(\phi,R) = R+\alpha R^2
 -8\pi G(\xi\phi^2R+g^{\mu\nu}\phi_{,\mu}\phi_{,\nu})$}

The dynamically equivalent Brans-Dicke action in $4$-dimensions was given
earlier in equation \ref{EX2ACTION},
where the Brans-Dicke field is required by the field equations to be
$G\Phi = 1 - 8\pi G\xi\phi^2 +2\alpha R$.
This can then be conformally transformed to the general relativistic
gravity lagrangian using the rescaled Einstein metric,
$\tilde{g}_{\mu\nu} = (G\Phi)g_{\mu\nu}$, with the corresponding matter
lagrangian containing the two scalar fields $\phi$ and
$\psi = \sqrt{\frac{3}{16\pi G}} \ln G\Phi$.
\begin{eqnarray}
\tilde{\cal L}_{\rm matter} & = & -\frac{1}{2}\tilde{g}^{\mu\nu} \left(
 \psi_{,\mu}\psi_{,\nu} + \exp(-\sqrt{16\pi G/3}\psi)
  \phi_{,\mu}\phi_{,\nu} \right) \nonumber \\
& & \ \  - V(\Phi,\phi) + \exp(-2\sqrt{16\pi G/3}\psi) {\cal L}_{\rm matter}
\end{eqnarray}
where the potential
\begin{equation}
V(\Phi,\phi) = \frac{1}{64\pi\alpha G}
 \left( 1- (1-8\pi G\xi\phi^2)\exp(-\sqrt{16\pi G/3}\psi) \right)^2
\end{equation}

\section{Summary}

I have expanded on the observation by Teyssandier and Tourrenc that the
classical field equations of fourth-order gravity theories may be derived
from an equivalent scalar-tensor lagrangian to show that this extends to
arbitrarily high-order theories.  Specifically, I have shown that the
field equations of ($4+2k$)th-order gravity can be derived from a
lagrangian where the Ricci scalar is coupled to a Brans-Dicke field with
$\omega=0$, but coupled through an interaction potential to a further
$k$ scalar fields.

Such scalar-tensor actions can be written in standard Einstein form by a
conformal re-scaling of the metric.  The gravitational actions written
in either frame are exactly equivalent, even including the correct boundary
terms. (The physical difference between the two metrics lies in the
matter lagrangian). By contrast there is no such conformal transformation of
the
original higher-order gravity action. One consequence is that there is no way
to write down the correct boundary term for the higher-order
theory\cite{MADSENBARROW89}
by conformally transforming the corresponding term in general relativity as one
can do in scalar-tensor gravity.

The field equations derived from higher-order lagrangians can be conformally
transformed to an Einstein frame with many scalar fields because of the
equivalence at the level of the classical field equations between these
theories and scalar-tensor gravity. I have referred to this as dynamical
equivalence. It is certainly sufficient to determine the classical
behaviour while the conformal factor is well-behaved, but it may not be valid
when considering ``off-shell'' or quantum effects where the field
configurations do not follow the classical trajectories.  In particular,
while calculations of quantum fluctuations during an inflationary epoch
may safely be done in the conformal Einstein frame of a scalar-tensor
theory\cite{KST90} one should beware of doing such calculations in
the conformal frame of higher-order gravity theories, where one is no
longer free to vary the scalar field $\psi$ independently of the metric.

It is instructive to note that some higher-order gravity theories cannot
be conformally transformed to an Einstein gravity with scalar fields, and
these correspond to those for which the dynamical equivalence with a
scalar-tensor theory breaks down.  For instance the gravitational
action
\begin{equation}
S = \frac{1}{16\pi G} \int_M d^4x \sqrt{-g}
 \left[ F_0(R,\phi) + F_1(R,\phi) \Box R
 +16\pi G \left( - \frac{1}{2}g^{\mu\nu}\phi_{,\mu}\phi_{,\nu} - V(\phi)
  \right) \right]
\end{equation}
can be transformed to a scalar-tensor model with $G\Phi=(\partial
F_0/\partial \varphi_0) + \varphi_1(\partial F_1/\partial\varphi_0) +
\Box F_1$, where $\varphi_0=R$ and $\varphi_1=\Box R$ along classical
trajectories, so long as the determinant of the matrix $F_{jk}$ (equation
\ref{Fjk}) is non-zero. This requires $\partial F_1/\partial\varphi_0 \neq 0$.
Amendola\cite{AMENDOLA93} recently
considered a gravitational lagrangian with a non-minimally derivative
coupled scalar field which differs only by a total divergence from the above
action with $F_1=F_1(\phi)$. Amendola was unable to find a conformal
Einstein frame because in this case the matrix $F_{jk}$ is degenerate.

\vspace{12pt}

I would like to thank John Barrow, Ed Copeland and Andrew Liddle for
many helpful comments. The author is supported by a SERC postgraduate
studentship and acknowledges use of the Starlink computer system at Sussex.



\end{document}